\documentclass[preprint,12pt]{elsarticle}
\usepackage{epsfig}
\usepackage{amssymb}

\usepackage{lineno,hyperref}
\modulolinenumbers[5]

\begin{document}
\begin{frontmatter}
\title{Relevant alternative analytic average magnetization calculation method for the square and the honeycomb Ising lattices}
\author[mymainaddress]{Tuncer Kaya}
\ead{tkaya@yildiz.edu.tr}
\address[mymainaddress]{ Department of Physics, Yildiz Technical University,
        34220 Davutpa\c sa, Istanbul, Turkey}

\begin{abstract}
In this work, the order parameter or average magnetization expressions are obtained for the square and the honeycomb lattices based on recently obtained magnetization relation,  $<\sigma_{0,i}>=
 <\!\!\tanh[ \kappa(\sigma_{1,i}+\sigma_{2,i}+\dots +\sigma_{z,i})+H]\!\!> $. Where, $\kappa$ is the coupling strength and $z$
is the number of nearest neighbors.  $\sigma_{0,i}$ denotes the
central spin at the $i^{th}$ site while $\sigma_{l,i}$,
$l=1,2,\dots,z$, are the nearest neighbor spins around the central
spin. In our investigation, inevitably we have to make a conjecture about the three site correlation function appearing in the obtained relation of this paper. The conjectured form of the the three spin correlation function is given by the relation, $<\!\!\sigma_{1}\sigma_{2}\sigma_{3}\!\!>=a<\sigma>+(1-a)<\sigma>^{(1+\beta^{-1})}$, here $\beta$ denotes the critical exponent for the average magnetization and $a$ is positive real number less than one. The relevance of this conjecture is based on fundamental physical reasoning. In addition, it is tested and investigated by comparing the obtained relations of this paper with the previously obtained exact results for the square and honeycomb lattices. It is seen that the agreements of the obtained average magnetization relations with those of the previously obtained exact results are unprecedentedly perfect.
\end{abstract}
\end{frontmatter}
%\linenumbers

\section{Introduction}
The Ising model, which was originated by Lenz in 1920 and was subsequently investigated by his student Ising in 1925 \cite{Ising}, unifies the study of phase transitions in systems as divers as ferromagnet, gas-liquids, binary alloys, and so on \cite{Stanley,Hu,Zhu,Keskin,Aouini}. In one-dimension (1D), it
has exact solution with the prediction of the absence of phase transition at finite temperature \cite{Huang, Baxter}. The Ising
model on the 2D lattices are the most popular since they exhibit not only the transition temperature \cite{book,Beale,McCoy} but also the free
energy and average magnetization relations for these lattices have been obtained exactly \cite{Yang83}. Exactly solvable
lattices in 2D, therefore, have special statuses for investigating
order-disorder phase transitions. They are important not only for
explicitly demonstrating the existence of phase transition at
finite temperatures but also because they established a benchmark
for simulations, motivated the research for exact solutions of other models and have served as a testing ground for the
efficiency of new theoretical ideas. At this point it is important to mention that the available exact expressions have been obtained with cumbersome transfer matrix method. Especially, exact calculation of the average magnetization relations require very challenging mathematical work. Therefore, there are only exact average magnetization expressions  for the 1D chain and for the 2D lattices. The method used to investigate the 2D lattices is to complicated to investigate the 3D lattices that any exact results still unknown. Indeed, we are not going to deal with the 3D lattices \cite{Perk1,Perk2} here in this paper, but we introduce a new approach which makes the mathematical calculation easily tractable so that it can be used in calculations and investigations of the other Ising lattices safely. In this paper we focus on, however, the honeycomb and square lattices to show the relevance of the approach of this paper and also for the sake of simplicity and also to test the validity and relevance of the approach which is going to introduced in this paper.

As the main purpose of this paper is going to be to calculate the average magnetization expressions for the square and honeycomb lattices, it is relevant to mention shortly the important achievements of the subject. The existence of the critical point was first proven by Kramers and Wanier \cite{Kramers} with the method of a dual transformation. The qualitative calculation of Kramers and Wanier was verified by Onsager \cite{16} and  \cite{Kaufman}. The existence of criticality  leaded to investigate and calculate the average magnetization. Although, the spontaneous magnetization of the Ising model on rectangular lattice was first calculated by Onsager \cite{Keh}, he never bothered or published his derivation. Yang was the first one who published his derivation of the average magnetization of the square lattice Ising model \cite{Yang1952}. And his result is $<\sigma>=[1-\sinh(2K)^{-4}]^{\frac{1}{8}}$. Since the method used in Yang's derivation is too cumbersome, he recall it as the longest calculation in his career  \cite{20}. Later, less complicated method, but are still to hard to recover, have been developed to obtain the partition function of the 2D Ising model \cite{26,27,28,33,34,Syozi}. If the other 2D Ising lattices are considered, an average magnetization relation was obtained by Naya for the Ising model on honeycomb lattice \cite{Naya} and by \cite{Potts} for the triangular lattice. Of course, there are other important contribution done on this subject, but it is impossible to mention all of them in this short paper. In other words, it goes beyond the scope of study.

For the sake of confirmation of the validity and relevance of the method going to be used in this paper, we are going to first consider to calculate an average magnetization expression for the honeycomb lattice since it produces less complicated interrelation between the order parameter and three spin correlation function.  In other words, we can investigate safely and readily the relevance  and importance of the method and proposed conjecture of this paper by studying the honeycomb lattice.  Second, the proposed method of this paper is going to be applied to the honeycomb structure. As it is going to be seen, the important step of this paper is to make a conjecture about the functional form of three spin correlation function. Therefore, it is also relevant to mention some previous studies related to the three spin correlation function. 

Three spin correlation function of the 2D Ising model was considered by Baxter \cite{114,Baxtery} for three spins surrounding a triangle. He used the Pfaffian method. Later, a simpler derivation was given by Enting \cite{Enting}, who also calculated the three spin correlation for the honeycomb lattice. There are also some other important studies on the subject of the three spin correlation functions \cite{104,105,106,Tanaka}  by the cited authors. The important consequence that we get from almost all of them is the common physical properties of the three spin correlation function: the three spin correlation function manifestly posses the same critical coupling strength and the same critical exponent as the order parameter. As we are going to see them in the next section, these physical properties are quite relevant and also apparently necessary to describe the three spin correlation function. Therefore, we are going to use these properties safely to propose a mathematical functional form for the three spin correlation function.

This paper is organized as follows. In the next section, the average magnetization of the honeycomb lattice  is going to be calculated with an alternative analytic method based on the previously derived average magnetization relation by us \cite{Kaya}. The obtained expression is also going to be compared by the already available exact result derived by Potts. In the third section, the average magnetization relation is going to be derived by the same method for the square lattice. And, the obtained result is going to compared with the exact result derived by Yang. In the same section, some discussions and conclusions are also going to be presented.

\section{The  average magnetization calculation of honeycomb lattice}

We start this section with the previously derived formula \cite{Kaya} which can be expressed in the absence of external magnetic field for the honeycomb lattice as
\begin{equation}
<\!\sigma_{0,i}\!>=
 <\tanh[ \kappa(\sigma_{1,i}+\sigma_{2,i}+\sigma_{3,i} )]> ,
\end{equation}
where $\kappa$ is the coupling strength and $\sigma_{0,i}$ denotes the
central spin at the $i^{th}$ site while $\sigma_{l,i}$,
$l=1,2,3$, are the nearest neighbor spins around the central
spin. At this point, it is important to mention that caring the index $i$ is not necessary from now on. Now, expressing the tangent hyperbolic function with the following equivalent relation as
\begin{equation}
\!\!\!\tanh[ \kappa(\sigma_{1}\!+\!\sigma_{2}\!+\!\sigma_{3}\! )]\!\!=\!\!A_{h}(\sigma_{1}+\sigma_{2}\!+\!\sigma_{3} \!)\!+\!B_{h}(\sigma_{1}\sigma_{2}\sigma_{3}).
\end{equation}
The parameter or the function $A$ and $B$ can be obtained quite easily as 
\begin{equation}
A_{h}=\frac{1}{4}\tanh(3K)+\frac{1}{4}\tanh(K)
;\ \ \ \   B_{h}=\frac{1}{4}\tanh(3K)-\frac{3}{4}\tanh(K).
\end{equation}
Substituting $A_{h}$ and $B_{h}$ into Eq. (11) and then Eq. (10), after a little bit algebra, Eq. (10) turns out to be
\begin{equation}
<\!\sigma\!>=\frac{\tanh(3K)-3\tanh K}{4-3(\tanh(3K)+\tanh K)}
 <\sigma_{1}\sigma_{2}\sigma_{3} > .
\end{equation}
\begin{figure*}[!hbt]
\begin{center}$
\begin{array}{cc}
\scalebox{0.7}{\includegraphics{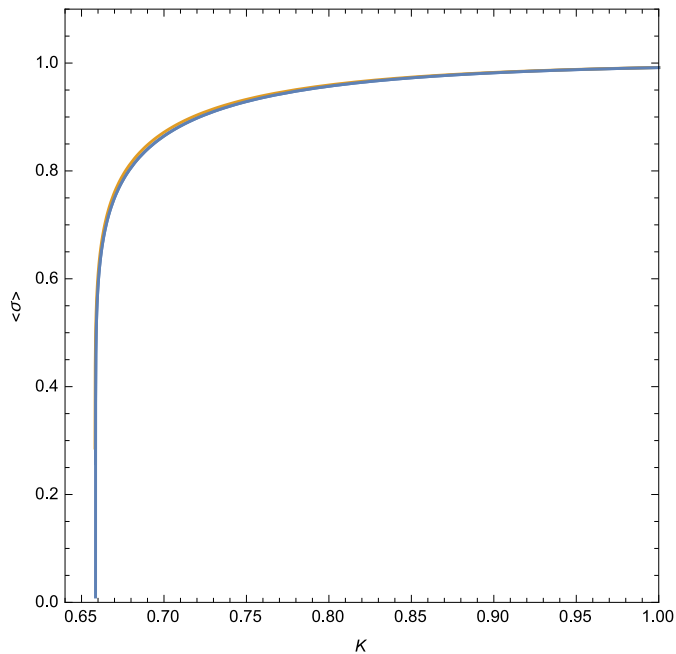}}
\end{array}$
\end{center}
\caption{The comparison of the average magnetization relation for the honeycomb lattice given in Eq. (7) with the exact expression given by Eq. (8). }
\end{figure*}
If the critical behavior of an order parameter is recalled, one can easily see that the three spins correlation function can be expressed as function of the order parameter. As pointed out before, it is not easy or not even possible to propose an exact relation for the three spins correlation, but taking into account general behavior of critical properties of the order parameter, one can conclude that the above conjectured functional form might be a proper one to use as an approximation for the three spins correlation unction. Therefore, the three spins correlation function can be constructed as $<\sigma_{1}\sigma_{2}\sigma_{3}>=a_{h}<\sigma>+(1-a_{s})<\sigma>^{1+\beta^{-1}}$, $\beta$ denotes the critical exponent of the 2D Ising system. Now, substituting this conjectured relation into the last equation, it leads to
\begin{equation}
<\sigma>=3A_{h}<\sigma>+B_{h}[a_{h}<\sigma>+(1-a_{h})<\sigma>^{1+\beta^{-1}}].
\end{equation}
The unknown parameter $a_{h}$ can be obtained if it is recalled that the order parameter vanishes for the values of $K$ less than the values of critical coupling strength, which is equal to $K_{h,c}=0.6585$ for honeycomb lattice. Thus, to calculate the value of $a_{h}$, the following condition can be obtained as
\begin{equation}
1-(3A_{h}+a_{h}B_{h})|_{K=K_{h,c}}=0.
\end{equation}
Solution of this equation produce $a_{h}=0.8034$.  It is important to notice that the same value for the parameter $a$ is also obtained by Baxter and Choy after a very cumbersome mathematical method \cite{Baxtery} and they called the parameter as "the critical amplitude". Remembering $\beta$ is equal to $\frac{1}{8}$ for the 2D Ising lattices, Eq. (14) leads to the following average magnetization relation for the honeycomb lattice for the values of $K>K_{h,c}$ 
\begin{equation}
<\!\sigma\!>=\Big{[}\frac{1-(3A_{h}+a_{h}B_{h})}{(1-a_{h})B_{h}}\Big{]}^{\frac{1}{8}}.
\end{equation}
The derived relation is compared with the exact result in Fig. 1  \cite{Codello}, given as
\begin{equation}
<\sigma>=\frac{(1+e^{4K}-4e^{4K})(1+e^{4K})}{(1+e^{2K})^{2}(1-e^{2K})^{6}}.
\end{equation}
As seen from the Fig. (1), the agreement of the obtained expression of this paper and the exact result is almost perfect. Even, it is almost impossible to see the differences in this figure. Therefore, we can claim that the obtained average magnetization relation given by Eq. (7) is almost exact. Meaning that  the conjecture used for the three spin correlation function is both relevant and valuable. Thus, we can use this approach for the calculation of the average magnetization for the other lattices. Indeed, in the following section we try to derive the average magnetization relation for the square lattice.

\section{The average magnetization derivation of square lattice}
In this section we apply the same method use in the previous section to calculate the average magnetization expression for the honeycomb lattice to to obtain a relation for the average magnetization of the square lattice. As it is done above, the procedure stars with writing the Eq. (1) for square lattice as
\begin{equation}
<\sigma_{0,i}>=
 <\tanh[ \kappa(\sigma_{1,i}+\sigma_{2,i}+\sigma_{3,i} +\sigma_{4,i})]>,
\end{equation}
where $\kappa$ is the coupling strength and $\sigma_{0,i}$ denotes the
central spin at the $i^{th}$ site while $\sigma_{l,i}$,
$l=1,2,3,4$, are the nearest neighbor spins around the central
spin. Expressing the tangent hyperbolic function with the following equivalent relation as 
\begin{equation}
\!\!\!\tanh[ \kappa(\sigma_{1}\!+\!\sigma_{2}\!+\!\sigma_{3}\! +\!\sigma_{4})]\!\!=\!\!A(\sigma_{1}+\sigma_{2}\!+\!\sigma_{3} \!+\!\sigma_{4})\!+\!B(\sigma_{1}\sigma_{2}\sigma_{3}+\sigma_{2}\sigma_{3}\sigma_{4}+\sigma_{3}\sigma_{4}\sigma_{1}+\sigma_{1}\sigma_{2}\sigma_{4}),
\end{equation}
where the functions $A$ and $B$ can be obtained quite readily as
\begin{equation}
A=\frac{1}{8}\tanh(4K)+\frac{1}{4}\tanh(2K)
;\ \ \ \   B=\frac{1}{8}\tanh(4K)-\frac{1}{4}\tanh(2K).
\end{equation}
Substituting these function into Eq. (1) leads to 
\begin{equation}
<\sigma>=<\tanh \kappa(\sigma_{1}\!+\!\sigma_{2}\!+\!\sigma_{3}\! +\!\sigma_{4})\!\!>=4A<\sigma>+4B<(\sigma_{1}\sigma_{2}\sigma_{3}>,
\end{equation}
where the simple facts $<\sigma>=<\sigma_1>=<\sigma_2>=<\sigma_3>=<\sigma_4>$ and \ \ \ $<\sigma_{1}\sigma_{2}\sigma_{3}>=<\sigma_{2}\sigma_{3}\sigma_{4}>=<\sigma_{3}\sigma_{4}\sigma_{1}>=<\sigma_{1}\sigma_{2}\sigma_{4}>$ are used.
Substituting the function into this equation, it can be written after some algebra as 
\begin{equation}
<\sigma>=\frac{\tanh4K-2\tanh2K}{2-\tanh4K-\tanh2K}<\sigma_{1}\sigma_{2}\sigma_{3}>.
\end{equation}
Taking into account the critical behavior of $<\sigma>$ which is zero for the coupling strength less than the corresponding critical coupling value $K_{c}$, it is easy to assume that the three spins correlation function $<\sigma_{1}\sigma_{2}\sigma_{3}>$ can be expressed as a function of order parameter $<\sigma>$. It is, of course, not easy to obtain the three spins correlation function in an exact form as a function of $<\sigma>$. But it is always possible to propose or conjecture up a particular form for it by taking into account the general properties of critical behavior. Now, we are ready to conjecture up the form of the three spins correlation function. For this purpose, it is proper to assume that the three spin correlation function can be expressed as in the form of $<\sigma_{1}\sigma_{2}\sigma_{3}>=a_{s}<\sigma>+b_{s}<\sigma>^{1+\beta^{-1}}$, $\beta$ denotes the critical exponent. Considering $<\sigma_{1}\sigma_{2}\sigma_{3}>=1$ as $K$ goes to infinity. One can easily see that $b_{s}=1-a_{s}$. Substituting this conjectured function into the
 Eq. (4), it turns out to be
\begin{equation}
<\sigma>=4A_{s}<\sigma>+4B_{s}[a_{s}<\sigma>+(1-a_{s})<\sigma>^{1+\beta^{-1}}].
\end{equation}
Now, using $<\sigma>=0$ for $K<K_{s,c}$, here $K_{s,c}=0.4407$ is the critical coupling strength for the square lattice, the parameter $a_s$ can be obtained from the following equation
\begin{equation}
1-4A_{s}(K_{s,c})-4B_{s}(K_{s,c})a_{s}=0
\end{equation}
as $a_{s}=0.7568$. Substituting the obtained values of $a_s$ and the values of the 2D critical exponent for magnetization, $\beta=\frac{1}{8}$, into
 Eq. (6), the following equation can be obtained for the values $K>K_{s,c}$
\begin{equation}
<\sigma>=\Big{[}\frac{1-4A_{s}-4B_{s}a_{s}}{4B_{s}(1-a_s)}\Big{]}^{\frac{1}{8}}.
\end{equation}
We are now in the position to investigate the relevance and accuracy of this relation. To this and, it might be proper to compare this relation with the exact relation obtained after very cumbersome calculation by Yang \cite{Yang1952} or Onsager, which is expressed as
\begin{equation}
<\sigma>=\Big{[}1-(\sinh2K)^{-4}\Big{]}^{\frac{1}{8}}.
\end{equation}
For comparison of the spontaneous magnetization expression obtained in this paper that of the exact result, they are plotted in Fig. 2. As seen from the figure they mach perfectly. Indicating that the obtained equation Eq. (16) is almost equivalent to the the exact result Eq. (17). Therefore, once again we can claim that the conjectured mathematical form for three spin correlation function is very relevant and valuable. In other words, the conjectured form can be use safely wherever it is necessary.
\begin{figure*}[!hbt]
\begin{center}$
\begin{array}{cc}
\scalebox{0.7}{\includegraphics{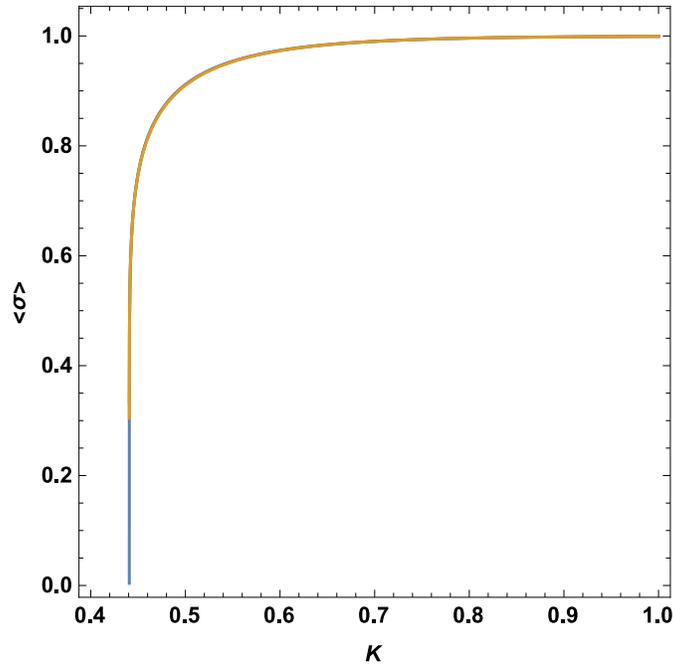}}
\end{array}$
\end{center}
\caption{The comparison of the obtained spontaneous magnetization expression given by Eq. (16) with the exact result obtained by Yang or Onsager.}
\end{figure*}

\section{Conclusion and Discussion} 
In this paper, a new approach is introduced to investigate and calculate the average magnetization of the honeycomb and the square lattices. To do so, we start our investigation by applying the published exact equation introduced previously. The application produces the relations which relate average magnetization of the lattices to the three spin correlation functions. Due to the mathematical form of the equation, it is concluded that the critical properties of the three spin correlation function has to be exactly the same as the critical properties of the average magnetization. Indeed, this point was also indicated 
by the other author previously. Taking into account these properties, the relation for the three spin correlation function is conjectured. The conjectured function has two parameters which can be obtained with some simple physical discussions. Indeed, these parameter are calculated quite easily. Substituting these parameters into the obtained relations lead to derive the average magnetization relations for the lattices. Comparing these obtained average magnetization expressions with those of the honeycomb and the square lattices, their relevance are tested. We conclude that the obtained spontaneous magnetization expressions of this paper are almost exactly the same as the previously obtained result of the honeycomb lattice by Naya and the obtained result of square lattice by Yang. In other words, in this paper, we obtain the average magnetization relations with very simple and tractable mathematical and physical approaches. I think that these points are important if we recall the mathematical procedure to obtain the cited exact results. More importantly, the method used in this paper may lead to a possible average magnetization treatment of 3D lattices such as the cubic lattice. In addition, our method may also possible leads to obtain the average magnetization relations of those lattices in the presence of external magnetic field. These points are going to be the subject of our future researches.

\end{document}